\documentclass[journal=apchd5,manuscript=article]{achemso}

\usepackage[T1]{fontenc}       % Use modern font encodings
\usepackage{inputenc}
\usepackage{color}
\usepackage{ulem}

\author{Maeliss Ethis de Corny}
\affiliation[UGA]
{Univ. Grenoble Alpes, Inst NEEL, F-38000 Grenoble, France}
\alsoaffiliation[CNRS]
{CNRS, Inst NEEL, F-38000 Grenoble, France}

\author{Nicolas Chauvet}
\affiliation[UGA]
{Univ. Grenoble Alpes, Inst NEEL, F-38000 Grenoble, France}
\alsoaffiliation[CNRS]
{CNRS, Inst NEEL, F-38000 Grenoble, France}

\author{Guillaume Laurent}
\affiliation[UGA]
{Univ. Grenoble Alpes, Inst NEEL, F-38000 Grenoble, France}
\alsoaffiliation[CNRS]
{CNRS, Inst NEEL, F-38000 Grenoble, France}

\author{Mathieu Jeannin}
\affiliation[UGA]
{Univ. Grenoble Alpes, Inst NEEL, F-38000 Grenoble, France}
\alsoaffiliation[CNRS]
{CNRS, Inst NEEL, F-38000 Grenoble, France}

\author{Logi Olgeirsson}
\affiliation[UGA]
{Univ. Grenoble Alpes, Inst NEEL, F-38000 Grenoble, France}
\alsoaffiliation[CNRS]
{CNRS, Inst NEEL, F-38000 Grenoble, France}

\author{Aur\'elien Drezet}
\affiliation[UGA]
{Univ. Grenoble Alpes, Inst NEEL, F-38000 Grenoble, France}
\alsoaffiliation[CNRS]
{CNRS, Inst NEEL, F-38000 Grenoble, France}

\author{Serge Huant}
\affiliation[UGA]
{Univ. Grenoble Alpes, Inst NEEL, F-38000 Grenoble, France}
\alsoaffiliation[CNRS]
{CNRS, Inst NEEL, F-38000 Grenoble, France}

\author{G\'eraldine Dantelle}
\affiliation[UGA]
{Univ. Grenoble Alpes, Inst NEEL, F-38000 Grenoble, France}
\alsoaffiliation[CNRS]
{CNRS, Inst NEEL, F-38000 Grenoble, France}

\author{Gilles Nogues}
\affiliation[UGA]
{Univ. Grenoble Alpes, Inst NEEL, F-38000 Grenoble, France}
\alsoaffiliation[CNRS]
{CNRS, Inst NEEL, F-38000 Grenoble, France}

\author{Guillaume Bachelier}
\affiliation[UGA]
{Univ. Grenoble Alpes, Inst NEEL, F-38000 Grenoble, France}
\alsoaffiliation[CNRS]
{CNRS, Inst NEEL, F-38000 Grenoble, France}
\email{guillaume.bachelier@neel.cnrs.fr}

\title{Wave-mixing origin and optimization in single and compact aluminum nanoantennas}

\keywords{Plasmonics, nonlinear optics, second harmonic generation, double resonance, mode matching, aluminum, single nanoantenna}

\begin{document}

\begin{tocentry}
\includegraphics[width=8.89cm]{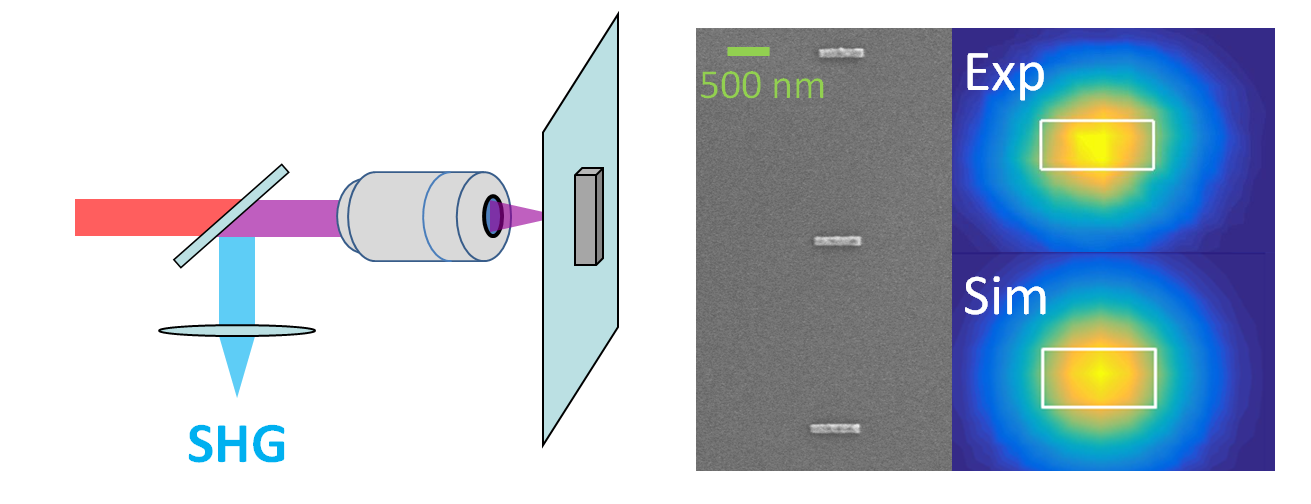}

\section*{For Table of Contents Use Only}

Wave-mixing origin and optimization in single and compact aluminum nanoantennas.\\

Maeliss Ethis de Corny, Nicolas Chauvet, Guillaume Laurent, Mathieu Jeannin, Logi Olgeirsson, Aur\'elien Drezet, Serge Huant, G\'eraldine Dantelle, Gilles Nogues, Guillaume Bachelier.\\

The present Table of Content (TOC) Graphic illustrates the setup (reflection mode) used to investigate the second harmonic generation (SHG) in single aluminum antennas ontained by e-beam lithography. The nonlinear signal recorded by scanning the sample under a tightly focused beam is quantitatively compared to numerical simulations taking into account the precise experimental configuration. 
\end{tocentry}

\begin{abstract}
  The outstanding optical properties for plasmon resonances in noble metal nanoparticles enable the observation of non-linear optical processes such as second-harmonic generation (SHG) at the nanoscale. Here, we investigate the SHG process in single rectangular aluminum nanoantennas and demonstrate that i) a doubly resonant regime can be achieved in very compact nanostructures, yielding a 7.5 enhancement compared to singly resonant structures and ii) the \(\chi_{\perp\perp\perp}\) local surface and \(\gamma_{bulk}\) nonlocal bulk contributions can be separated while imaging resonant nanostructures excited by a tightly focused beam, provided the \(\chi_{\perp\parallel\parallel}\) local surface is assumed to be zero, as it is the case in all existing models for metals. Thanks to the quantitative agreement between experimental and simulated far-field SHG maps, taking into account the real experimental configuration (focusing and substrate), we identify the physical origin of the SHG in aluminum nanoantennas as arising mainly from \(\chi_{\perp\perp\perp}\) local surface sources.
\end{abstract}

\section*{Introduction}
Controlling and amplifying nonlinear processes at the nanometer scale is a long-standing quest for practical applications including quantum telecommunication and quantum computing. The greatest challenge is to overcome the intrinsically low efficiency of nonlinear materials, which is further reduced upon down-sizing the active medium to the nanoscale. To this end, plasmonic nanostructures have been proposed as subwavelength resonators for enhancing nonlinear processes such as second (SHG) and third (THG) harmonic generations\cite{Hanke2009,Metzger2012,Aouani2012,Ginzburg2012,Hentschel2012}. The main advantage of these structures lies in the tunability of their resonances as a function of size, morphology or coupling configuration\cite{Noguez2007,Hecht2012,Novotny2011,Maier2005}. They also strongly benefit from their ability to confine the electromagnetic field below the diffraction limit, boosting their nonlinear efficiency accordingly thanks to the reduced mode volumes\cite{Kauranen2012,Canfield2007}. In this context, doubly resonant structures are very promising since they benefit from the field enhancement at both excitation and emission steps\cite{Thyagarajan2012}. Yet, for optimized frequency conversion yields, these plasmonic resonances must be further spatially mode-matched. In other words, the corresponding amplitude and phase distributions of the electric fields have to be tailored in such a way that the generated near-field nonlinear currents coherently radiate their harmonic photons in the far-field. Such structures have been recently achieved by coupling a gold nanorod to a V-shaped antenna with typical sizes down to 150 nm\cite{Celebrano2015}. Here, we demonstrate that these double-resonance and mode-matching conditions can be reached in a single and compact aluminum antenna for application purposes in the visible range\cite{Knight2014,Knight2012,Metzger2015}. The net advantage compared to gold-based nanostructures is the lack of superimposed multiphoton luminescence background, allowing a direct read-out of pure SHG signals. Thanks to this specific property, we show how the origin of the SHG response can be unambiguously unveiled by quantitatively comparing measured and simulated maps of the nonlinear signals, which are obtained by imaging a single antenna under a tightly focused beam.

\section*{Results and discussion}

In order to infer the origin of the nonlinear response of single aluminum antennas, we first investigated their spectral response at a fixed excitation wavelength of 850 nm by scanning the monochromator from 400 to 650 nm (see the experimental section for more details). Only one signal peak is observed at 425 nm, i.e. at the harmonic wavelength. Its intensity is further shown to scale as the squared optical excitation power:  \(I(2\omega)\propto[I(\omega)]^2\) with a correlation coefficient equal to \(r^2=0.996\), unambiguously assigning its origin to a SHG process. This result significantly differs from the case of gold nanoantennas (fabricated by the same process) where the sharp peak at the harmonic wavelength is superimposed to a broadband signal with a maximum intensity at 550 nm (see Figure 1). This response is known to originate from the two-photon luminescence (TPL) of gold, which is related to its band structure close to the \(X\) and \(L\) points providing large densities of states\cite{Dionne2013,Wang2014,Viarbitskaya2013,Biagioni2012,Biagioni2009}. Even higher order photoluminescence has been reported in the literature, smearing the weak SHG signal in an intense and incoherent background\cite{Deng2013}. Here, the background-free signal of aluminum can be directly exploited and quantitatively compared to simulations.

\begin{figure*}[h!t]
\centering
\includegraphics[width=10cm]{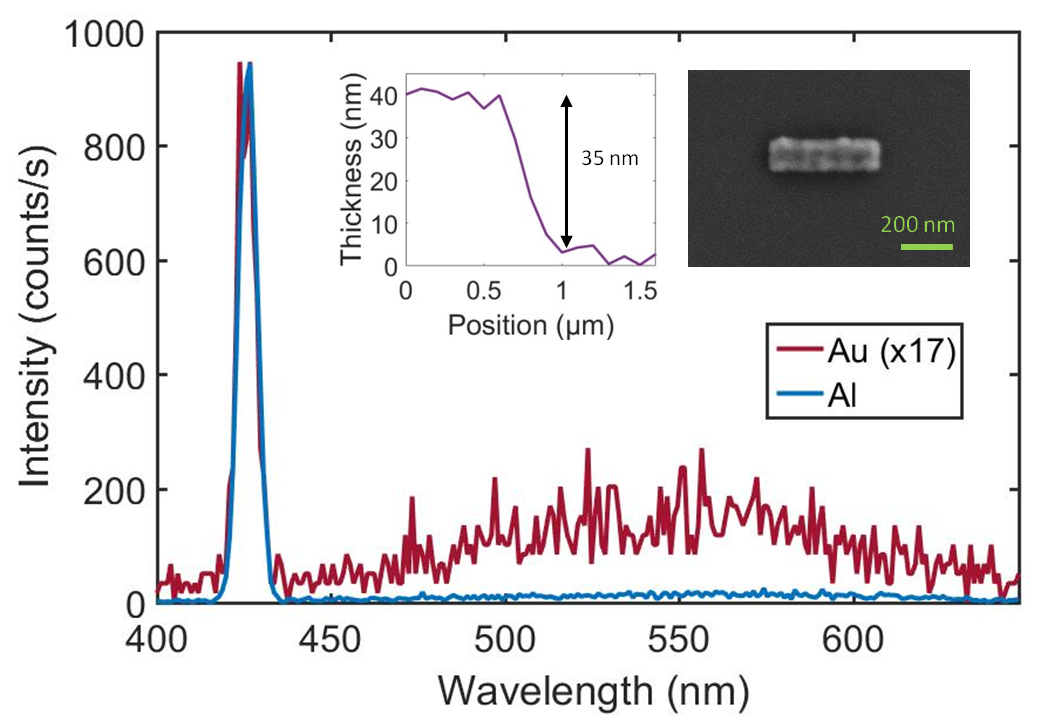}
	
  \caption{SHG intensity spectrum of single 225 nm-long aluminum (blue) and gold (red) antennas illuminated at 850 nm. In inset, the SEM image of a single 425 nm aluminium nanoantenna (right) and typical topographic profile measured with the probe of a scanning near-field optical microscope implemented in the setup (left).}
  \label{Fig1}
\end{figure*}

Another salient feature of aluminum antennas is their ability to sustain plasmonic resonances over the entire visible spectrum, in contrast to gold nanostructures where the resonances are hampered above the interband transition threshold (near 520 nm) due to the strong absorption rate arising from the generation of electron-hole pairs\cite{Knight2014,Knight2012}. This allows promoting the concept of both double-resonance and mode-matching conditions\cite{Celebrano2015} in a single and compact aluminum nanoantenna. To evidence this peculiar property, absorption spectra were simulated by placing in the vicinity of the nanostructure a point-dipole source aligned with the long antenna axis. This allows exciting all possible plasmon resonances associated either with bright or dark modes in the nanostructures\cite{Nogues2013}. As shown in Figure 2 for 225 nm and 425 nm-long antennas, the spectral distribution of the plasmonic resonances can be tailored by adjusting the antenna aspect ratio. It is therefore possible to reach resonant conditions at the excitation wavelength, at the emission wavelength, or at both wavelengths, depending on the targeted application or investigation as shown later. 

\begin{figure*}[h!t]
\centering
\includegraphics[width=14cm]{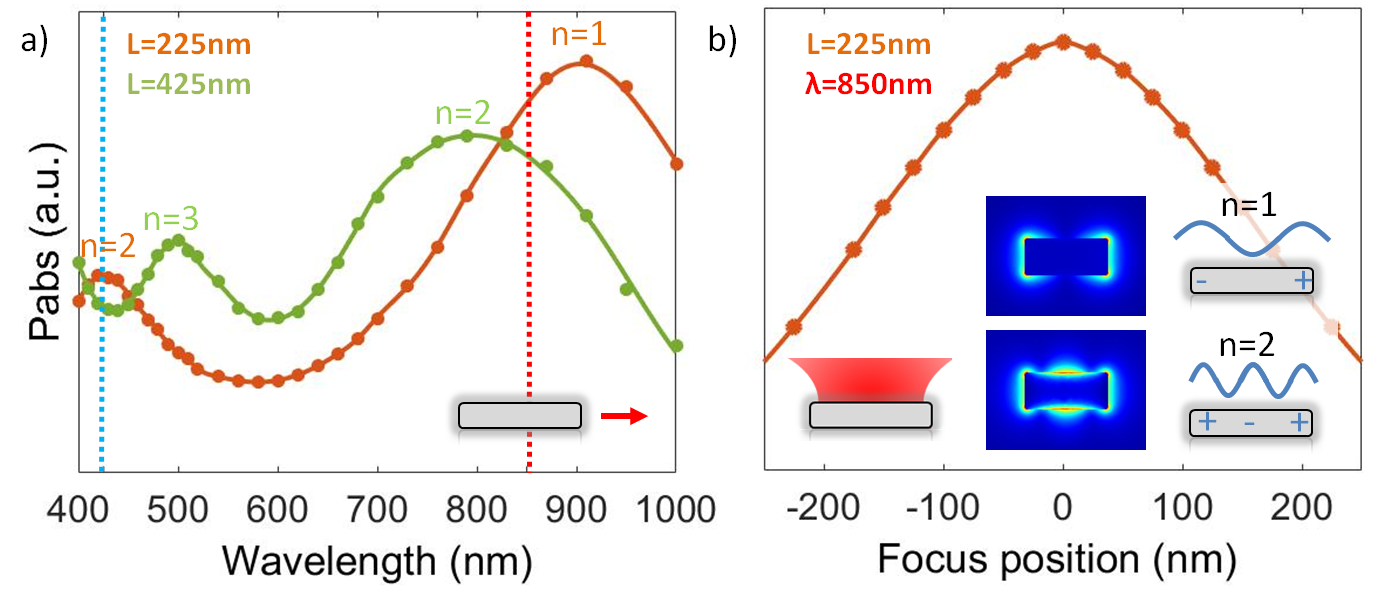}
	
  \caption{a) Simulated absorption spectra for a 225 nm (orange curve) and a 425 nm (green curve) aluminum nanoantennas excited by an electric point dipole in its vicinity. Plasmonic mode orders n are indicated over each resonance. Red and blue dashed lines indicate excitation and detection wavelengths used for all measurements, respectively. b) 
Simulated absorption spectra for a 225 nm scanned along its long axis under a focused beam at the fundamental frequency (850 nm). Inset: 2D maps of the scattered electric field norm at the fundamental and harmonic frequencies obtained inside the antenna when the focus of the microscope objective is centered on the antenna. They evidence the \(n=1\) fundamental and \(n=2\) excited plasmon modes, respectively. The excitation configurations for the absorption spectra are sketched by a) the red dipole and b) focused beam, respectively.}
  \label{Fig2}
\end{figure*}

Even more interesting is the ability to choose the symmetry of the plasmon modes for the excitation and the emission steps in order to maximize the mode matching at the nanoscale. To give a simple picture, let us consider the dipolar resonance that is the most efficiently excited localized surface plasmon for subwavelength nanostructures. It is associated with enhanced electric field amplitudes at each antenna apex with opposite surface charges (see Figure 2). Assuming the SHG currents to be proportional to the squared electric field (this will be discussed in detail below), their spatial distribution has here a symmetric character with two lobes sharing the same amplitude and phase at the antenna apexes. The plasmon mode matching this condition at the emission step is the quadripolar plasmon shown in Figure 2. More generally, a perfect mode matching with the \(n^{th}\) plasmon resonance excited at the fundamental frequency will be obtained with the \(2n^{th}\) mode having an opposite parity at the harmonic frequency. These plasmon modes are often referred to as ``dark'' modes since they hardly couple to plane waves propagating perpendicularly to the substrate. However, they do emit light at large angles so that harmonic photons are efficiently collected by large NA objectives. 

\begin{figure*}[h!t]
\centering
\includegraphics[width=16cm]{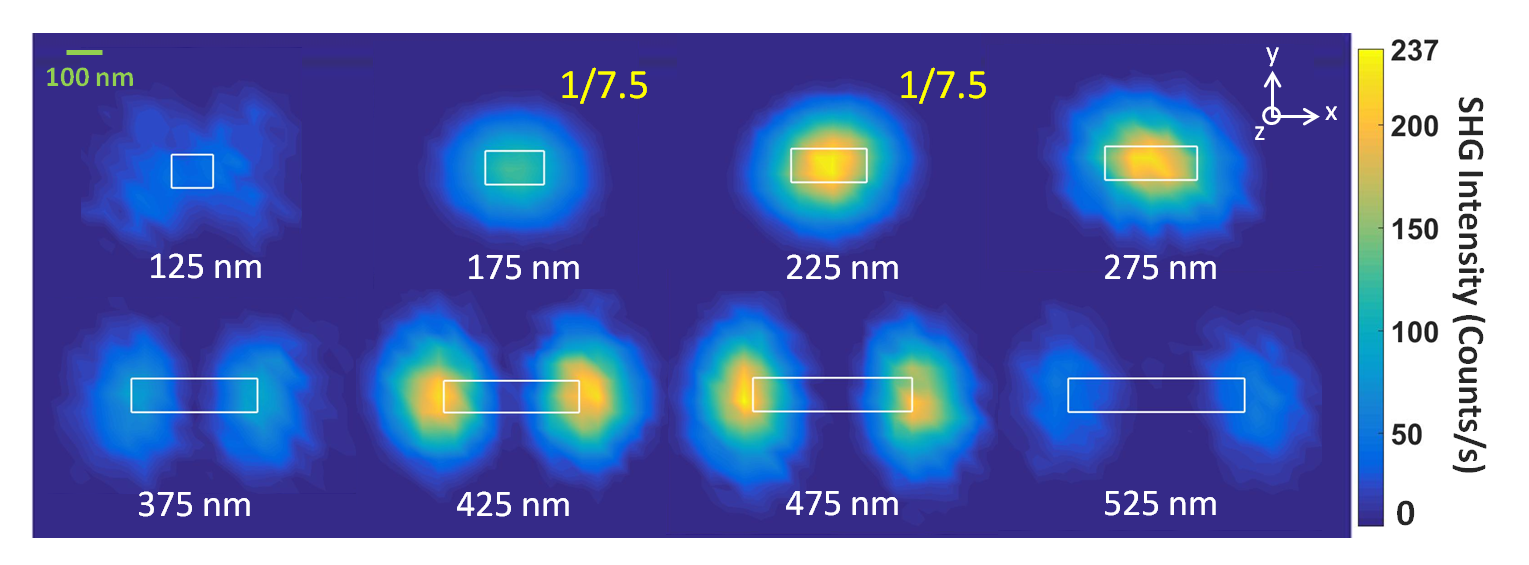}
  \caption{SHG excitation intensity maps in counts/s excited at 850 nm for aluminum nanoantennas of increasing  lengths from 125 to 525 nm. White rectangles are guides to the eyes indicating the position of corresponding nanostructures. Both excitation and detection linear polarizations are set along the x axis.}
  \label{Fig3}
\end{figure*}

In order to demonstrate these two major properties -- double resonance and mode matching conditions -- 2D maps of the SHG intensity scattered in the far-field were recorded by scanning the sample (x-y plane) under the focused laser beam (z axis), see Figure 3. The first conclusion that can be easily drawn from these measurements is that the spatial distribution of the SHG intensity depends on the antenna length, as expected from the tunability of the plasmon resonance. However, interpreting the SHG data is not straightforward here since the nanostructure is lying on a substrate, is excited by a strongly focused beam and, more importantly, is recorded while scanning the sample. This means that we do not image a given plasmon mode emitting the SHG signal but rather the overall efficiency of harmonic photon generation and collection as a function of the focus point position. For example, the 225 nm antenna exhibits only one lobe in contrast with the two-lobe picture associated with the excited dipolar plasmon mode (see inset of Figure 2). Therefore, near-field properties and far-field measurements associated with sample scanning in reflection mode must be disentangled: the dipolar mode is maximally excited at the center of the 225 nm antenna, as shown in Figure 2b by the absorption curve obtained by scanning the nanoantenna along its long axis, despite the fact that it shows enhanced electric fields at the antenna endings. Yet, the diffraction-limited resolution provided by the immersion objective allows resolving and differentiating the spatial SHG distribution associated with each single antenna (one or two lobes as shown in Figure 3).

The second outcome of Figure 3 is related to the SHG enhancement by plasmonic resonances. While non-resonant antennas generate signals as low as 50 photons/s, 425 nm-long antennas reach roughly 250 photons/s, largely surpassed by the 225 nm-long antennas with typically 1800 photons/s in the same experimental conditions. In terms of magnification, this corresponds to a 5-fold enhancement from non resonant to simply resonant (at the excitation step) antennas and up to a 36-fold enhancement in the doubly resonant regime. This double resonance condition can also be inferred from simulated near-field maps such as those shown in Figure 2 for the 225 nm-long antenna: a pure dipole and a pure quadrupole are obtained at the fundamental and harmonic frequency respectively, whatever the antenna position with respect to the focused beam (data not shown due to their similarity with those of Figure 2). This is a complementary proof that the plasmon modes are resonantly excited, since otherwise a coherent superposition of modes having distinct node numbers would have been observed. Note that we did not attempt to perfectly optimize the nanostructure morphology so that the net enhancement might be even larger, demonstrating that compact aluminum antennas are ideal candidate for efficient and compact SHG boosters.

Accounting for the complex readout of the optical maps (the far-field excitation and the collection locations are changed while scanning the sample) requires a complete simulation of the experimental configuration. First, the focusing and collection by optical elements is treated analytically (Matlab\textregistered) by wavelet sums\cite{Novotny2012}. This allows accounting for standard lenses as well as for large NA objectives, the latter producing longitudinal electric field components that are fully taken into account beyond the paraxial approximation. Second, the effect of immersion oil and substrate are described in the wavevector space through Fresnel transmission and reflection coefficients\cite{Novotny2012}. Third, a full numerical simulation (FEM, Comsol\textregistered 4.3a) is performed at each position of the experimental 2D maps, from the excitation and collection processes to the integration over the detector surface\cite{Gloppe2014}. Therefore, due to the large computation resource requirements, the simulations have been implemented on a home-made cluster in order to allow parametric investigation as a function of wavelength, polarization, particle morphology etc. 

The treatment of the SHG response can be found in Ref.~\citenum{Bachelier2008,Butet2010}, although we adopt here a different strategy to establish the origin of the nonlinearity. Instead of analyzing the far-field radiation diagrams as a function of the incident polarization, we investigate the 2D cartography of the SHG intensity distribution over a single nanostructure. Hence, for each recorded map, two simulations have been performed in order to account for two distinct contributions separately: the breakdown of the centrosymmetry at the metal surface and the intensity gradients inside the bulk material\cite{Bachelier2010}. The corresponding local surface and nonlocal bulk polarizations, \(\mathbf{P}_{surf,\perp}\) and \(\mathbf{P}_{bulk}\)  respectively, are given by: 
\begin{equation}
\label{Psurf}
\mathbf{P}_{surf,\perp}(\mathbf{r},2\omega)=\chi_{\perp\perp\perp}E_{\perp}(\mathbf{r},\omega)E_{\perp}(\mathbf{r},\omega)\mathbf{n}
\end{equation}
\begin{equation}
\label{Pbulk}
\mathbf{P}_{bulk}(\mathbf{r},2\omega)=\gamma_{bulk}\mathbf{\nabla}.\left[\mathbf{E}(\mathbf{r},\omega).\mathbf{E}(\mathbf{r},\omega)\right],
\end{equation}
where the electric fields \(\mathbf{E}(\mathbf{r},\omega)\) are evaluated inside the metal and the polarization vector \(\mathbf{P}_{surf,\perp}\) lead to surface currents located just outside the metal \cite{Sipe1980}. The subscript \(\perp\) indicates a component along the unit vector \(\mathbf{n}\), perpendicular to the surface and pointing out of the metal. Note also that the bulk polarization \(\mathbf{P}_{bulk}\) is entirely evaluated inside the particle; it does not act across the particle surface and therefore does not contribute in this description to the nonlinear surface current. Following Sipe et al.\cite{Sipe1980}, the second order susceptibility tensor elements associated with local surface and nonlocal bulk currents, \(\chi_{\perp\perp\perp}\) and \(\gamma_{bulk}\) respectively, are defined in our model by\cite{Bachelier2010}:
\begin{equation}
\chi_{\perp\perp\perp}=-\frac{a}{4}\left[\epsilon_r(\omega)-1\right]\frac{e\epsilon_0}{m\omega^2}
\end{equation}
\begin{equation}
\gamma_{bulk}=-\frac{d}{8}\left[\epsilon_r(\omega)-1\right]\frac{e\epsilon_0}{m\omega^2}
\end{equation}
where \(m\) and \(e\) are the electron mass and charge and \(\epsilon_0\epsilon_r(\omega)\) is the dielectric function of the metal\cite{Rakic1998} depending on the pulsation \(\omega\). \(a\) and \(d\) are the adimensional Rudnick and Stern parameters\cite{Rudnick1971}, equal to 1 and -1 respectively in the hydrodynamic model applied to the conduction electrons\cite{Sipe1980}. With these expressions and taking into account the real experimental constraints, the SHG simulated intensities are expressed in terms of photons/s and can be directly compared to measurements. 

As discussed in detail in Ref.~\citenum{Wang2009}, there are several other contributions coming either from the particle surface, \(\chi_{\perp\parallel\parallel}\) and \(\chi_{\parallel\parallel\perp}\), or from the bulk \(\beta_{bulk}\mathbf{E}(\mathbf{r},\omega)\left[\mathbf{\nabla}.\mathbf{E}(\mathbf{r},\omega)\right]\) and \(\delta'_{bulk}.\left[\mathbf{E}(\mathbf{r},\omega).\mathbf{\nabla}\right]\mathbf{E}(\mathbf{r},\omega)\). According to their measurements on gold planar substrates, their associated nonlinear susceptibilities are at least one order of magnitude smaller than those considered in Eqs.~\ref{Psurf} and~\ref{Pbulk}, leading to SHG signals more than two orders of magnitude weaker than the leading contributions investigated here. The SHG intensities measured at the single particle level being of the order of a few hundreds of photons/s, they do not allow extracting these extra contributions lying below the shot noise in our experiments.

\begin{figure*}[h!t]
\centering
\includegraphics[width=16cm]{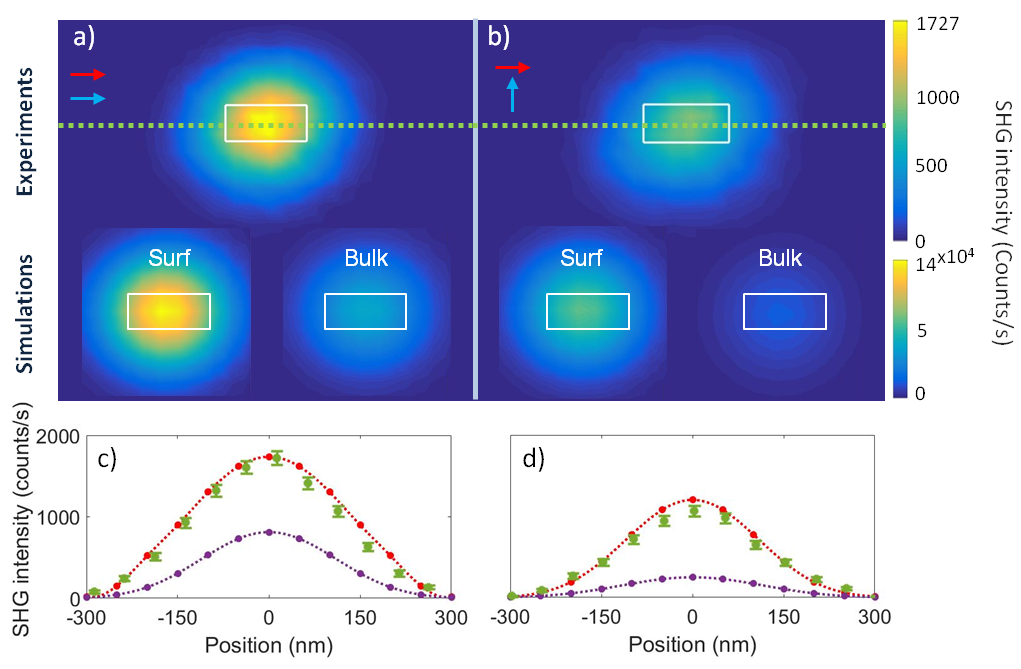}
  \caption{Measured and simulated SHG maps obtained for a single 225 nm aluminum antenna excited at 850 nm with an electric field parallel to the antenna axis (red arrows). The detection polarization is changed between panels (a) and (b) as indicated by the blue arrows. The SHG maps simulated with \(\chi_{\perp\perp\perp}\) local surface and \(\gamma_{bulk}\) nonlocal bulk contributions are labeled Surf and Bulk, respectively. The intensity profiles along the dashed green lines in panels (a) and (b) are shown in panels (c) and (d). The simulated curves for the \(\chi_{\perp\perp\perp}\) local surface (red dashed line) and \(\gamma_{bulk}\) nonlocal bulk (violet dashed line) contributions are divided by 79 (see text) to match the experimental data in panel (c).}
  \label{Fig4}
\end{figure*}

The doubly resonant 225 nm-long antennas were investigated first (see Figure 4) with an incident polarization parallel to their long axis. Both detection polarizations are measured simultaneously on two distinct single-photon counting modules (SPCMs) in order to ensure the exact position correlation between the two maps. As evidenced in Figures 4(a) and 4(b), the measured and simulated SHG maps show a good agreement in terms of intensity distribution and polarization responses. The mismatch between the measured and calculated signal levels are due to the optical element transmissions and the sensitivity of the SPCMs. This was checked by calibrating our setup along the collection path (27.3\% transmission) and by taking into account the datasheets of the microscope objective (80\% transmission) and of the SPCMs (\~25\% efficiency at 450 nm). Further considering the actual values of the Rudnick and Stern parameters for aluminum (see below) leads to an expected ratio of approximately 70 between simulations and experiments, to be compared to the actual value of 79 obtained by dividing the respective signals expressed in counts/s. To be more quantitative, the measured and calculated intensity profiles along the long axis of the antenna are shown in Figure 4 (c) and (d). The corresponding widths and polarization ratios are in perfect agreement up to the experimental uncertainties corresponding to the shot noise (see green error bars): in both cases, the SHG signal collected for a polarization parallel to the antenna long axis is 1.5 times higher than for crossed polarizations.

More interesting is the comparison between the relative efficiencies of the \(\chi_{\perp\perp\perp}\) local surface and \(\gamma_{bulk}\) non-local bulk contributions obtained here using unitary Rudnick and Stern parameters of the hydrodynamic and labeled accordingly in Figure 4. In contrast to an intuitive belief, both contributions give the same intensity profile centered on the antenna. First, as the fundamental electric field is maximal at the antenna apexes (dipolar resonance) and since the \(\chi_{\perp\perp\perp}\) local surface contribution is proportional to the electric-field squared, one might have expected one lobe on each side of the antenna. As evidenced by the white rectangles mimicking the antennas, this is clearly not the case here, even taking into account the diffraction-limited spots. Second, the intensity profiles are the same for the \(\chi_{\perp\perp\perp}\) local surface and \(\gamma_{bulk}\) non-local bulk contributions despite the fact that the nonlinear sources have neither the same locations nor the same intensity dependence (remember that the \(\gamma_{bulk}\) non-local bulk current is proportional to the gradient of the electric field squared). This is apparently in agreement with their inseparability for flat surfaces as discussed in Ref.~\citenum{Wang2009}.

\begin{figure*}[h!t]
\centering
\includegraphics[width=16cm]{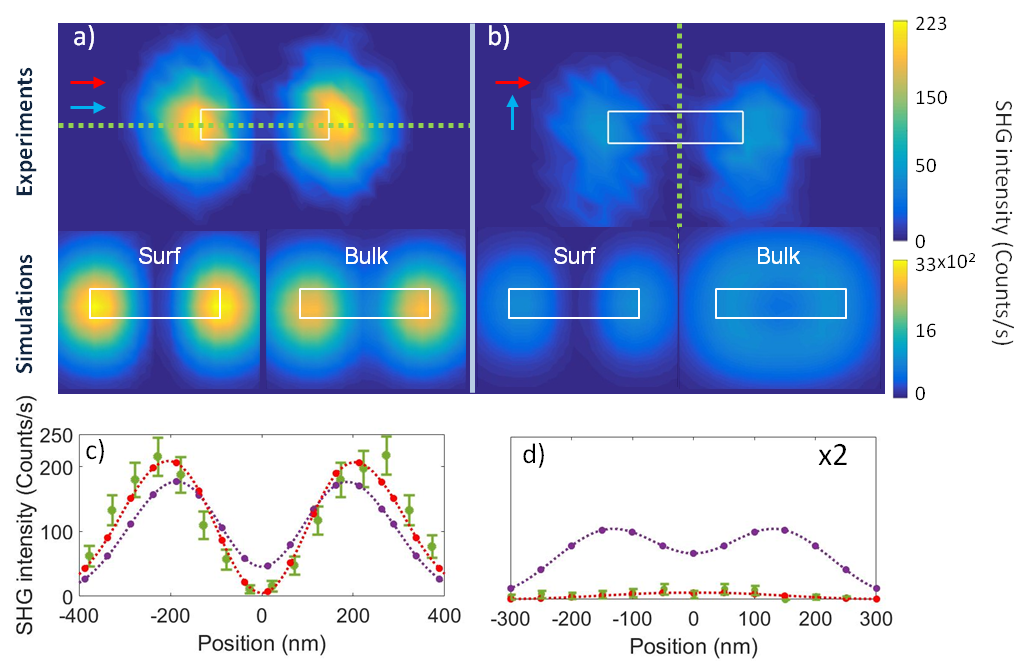}
  \caption{Same as in Figure 4 for a single 425 nm antenna. Note that the intensity profiles of panel (d) have been obtained perpendicularly to the antenna long axis, as shown by the green dashed line in panel (b), and multiplied by a factor of two for a better visibility.}
  \label{Fig5}
\end{figure*}

As discussed in Ref.~\citenum{Sipe1987}, the \(\gamma_{bulk}\) nonlocal contribution can be included in an effective surface contribution, so that the normal component of the polarization sheet reads:
\begin{equation}
P_{eff,\perp} (\mathbf{r},2\omega)=\chi_{\perp\parallel\parallel} E_{\parallel} (\mathbf{r},\omega)E_{\parallel}(\mathbf{r},\omega)+\chi_{\perp\perp\perp} E_{\perp}(\mathbf{r},\omega) E_{\perp}(\mathbf{r},\omega)+\frac{\gamma_{bulk}}{\epsilon(2\omega)} E(\mathbf{r},\omega)E(\mathbf{r},\omega)
\end{equation}
which can be recast into,\cite{Sipe1987}
\begin{equation}
P_{eff,\perp} (\mathbf{r},2\omega)=\left[\chi_{\perp\perp\perp}-\chi_{\perp\parallel\parallel}\right] E_{\perp} (\mathbf{r},\omega)E_{\perp}(\mathbf{r},\omega)+ \left[\chi_{\perp\parallel\parallel}+\frac{\gamma_{bulk}}{\epsilon(2\omega)}\right] E(\mathbf{r},\omega)E(\mathbf{r},\omega)
\end{equation}
This expression has been used to show that the only separable quantities are \(\chi_{\perp\perp\perp}-\chi_{\perp\parallel\parallel}\) and \(\chi_{\perp\parallel\parallel}+\gamma_{bulk}/\epsilon(2\omega)\), or in other word that the three quantities \(\chi_{\perp\perp\perp}\), \(\chi_{\perp\parallel\parallel}\) and \(\gamma_{bulk}\) cannot be extracted independently from optical experiments in the far-field. Yet, as discussed in Ref.~\citenum{Teplin2002}, the existing models for the nonlinear response of metals predict that \(\chi_{\perp\parallel\parallel}=0\), so that the polarization reduces to
\begin{equation}
P_{eff,\perp} (\mathbf{r},2\omega)=\chi_{\perp\perp\perp} E_{\perp}(\mathbf{r},\omega) E_{\perp}(\mathbf{r},\omega)+\frac{\gamma_{bulk}}{\epsilon(2\omega)} E(\mathbf{r},\omega)E(\mathbf{r},\omega).
\end{equation}

As the two nonlinear currents are not linked to the electric field in the same way, spatial decorrelation between SHG intensity distributions for the \(\chi_{\perp\perp\perp}\) local and the \(\gamma_{bulk}\) non-local responses should occur at least in specific configurations, allowing weighing their relative contribution by comparing measured and simulated intensity profiles. We have therefore numerically investigated the SHG responses associated to \(\chi_{\perp\perp\perp}\) local and \(\gamma_{bulk}\) non-local contributions for various antenna aspect ratios in order to find parameters where the spatial decorrelation is pronounced and experimentally addressable. As shown in Figure 5(b), this is achieved for 425 nm-long antennas when the detection polarization is perpendicular to the antenna long axis, leading to different spatial patterns (see the subpanel corresponding to the simulations). Hence, our simulations unambiguously demonstrate that the \(\chi_{\perp\perp\perp}\) local surface and the \(\gamma_{bulk}\) non-local bulk contributions are separable while imaging resonant and nano-sized structures excited by a tightly focused beam. Therefore, they are not mergeable into a single effective nonlinear surface contribution as it is the case for planar surfaces illuminated by planar waves\cite{Wang2009}. Note that the \(\chi_{\perp\parallel\parallel}\) local surface contribution would also produce a SHG intensity map distinct from those of \(\chi_{\perp\perp\perp}\) and \(\gamma_{bulk}\) in the crossed polarization configuration (data not shown). It is thus mergeable with none of them. However, if \(\chi_{\perp\parallel\parallel}\) was not zero, it would be possible to reproduce the SHG intensity distribution associated with the \(\gamma_{bulk}\) nonlocal bulk contribution by combining the \(\chi_{\perp\parallel\parallel}\) and \(\chi_{\perp\perp\perp}\) local surface sources, as expected from Equation~6 using the appropriate values for the nonlinear susceptibilities.

Despite the weak measured signal in the crossed polarization configuration, the SHG intensity pattern clearly matches the \(\chi_{\perp\perp\perp}\) local surface contribution and departs from the \(\gamma_{bulk}\) nonlocal bulk one, the latter leading a donut-like distribution in the simulated maps. This suggests that the \(\chi_{\perp\perp\perp}\) local surface current is the leading SHG source in Aluminum. In order to support this statement, intensity profiles were extracted from the measurements and simulations along two crossed directions (see the green dashed lines in Figure 5, different from the ones of Figure 4). Several insights into the nonlinear behaviors can be gained. First, the signal ratio between \(\chi_{\perp\perp\perp}\) local surface and \(\gamma_{bulk}\) nonlocal bulk contributions does depend on the antenna length, i.e. on the actual field distributions and gradients or the plasmon resonances involved. Generally speaking, it is therefore not straightforward, nor even correct, to assign a leading origin to the nonlinear response as the relative strength of \(\chi_{\perp\perp\perp}\) local surface and \(\gamma_{bulk}\) nonlocal bulk contributions is strongly affected by the particle morphology and the excitation/collection configuration. Yet, as far as the description in terms of Rudnick and Stern parameters is relevant, these coefficients entering in the simulations are constant from one structure to another and extractable from the experimental data as done in Ref.~\citenum{Bachelier2010}. Here, it is performed at the single nanostructure level and taking into account the effect of focusing and collection by the microscope objective together with the substrate effect. A close inspection of the simulated and measured intensity profiles along the antenna, see Figure 5(c), shows that the experimental data and the simulated signals for the \(\chi_{\perp\perp\perp}\) local surface contribution nearly vanish at the center of the aluminum antenna, in contrast to the \(\gamma_{bulk}\) nonlocal bulk contribution. This implies that the latter is largely overestimated with \(d=-1\). This conclusion is further confirmed in the crossed polarization configuration shown in Figure 5(b), where the \(\gamma_{bulk}\) nonlocal bulk contribution leads to a donut-like intensity distribution drastically different from the experimental data. It is even clearer with the intensity profiles taken perpendicular to the antenna in Figure 5(d), where the measured data perfectly follows the \(\chi_{\perp\perp\perp}\) local surface contributions, whereas the \(\gamma_{bulk}\) nonlocal bulk response is by far too large (at least by a factor 6.5) and shows a double bell curve that is not observed experimentally. This discrepancy with the hydrodynamic model has already been pointed out for the second-order nonlinear susceptibility of thin aluminum films\cite{Teplin2002}. Taking into account the difference in notations (a factor of 2 compared to Eq.~4), the Rudnick and Stern parameters were measured to be \(|a|=2.3\pm 0.72\) and \(|d|=0.018\pm 0.004\) . If the \(\chi_{\perp\perp\perp}\) local surface contribution is comparable to that of gold\cite{Bachelier2010}, the \(\gamma_{bulk}\) nonlocal bulk current is relatively 250 times smaller leading to a vanishingly small scattered intensity. We have also evaluated the \(\chi_{\parallel\parallel\perp}\) local surface contribution using the Rudnick and Stern parameter \(|b|=0.0146 \pm 0.004\) of Ref.~\citenum{Teplin2002} (data not shown) and obtained less than 0.1 photons/s, far below the shot noise level. Furthermore, neither the polarization ratio of the simulated maps nor the intensity in the center of the antenna is compatible with the measured data of Figure 5, so that this contribution can be ruled out in the present experiments.
This clearly supports our findings for single aluminum nanostructures, although no clear theoretical explanation has been found to account for this drastic deviation from the hydrodynamic model\cite{Teplin2002}.

\section*{Conclusion}
In conclusion, we have demonstrated that doubly resonant and mode matching conditions can be achieved for SHG with a single and compact aluminum nanoantenna by adjusting its morphological parameters. Substantial SHG enhancements are obtained compared to simply resonant structures (typical 8-fold magnification) and to nonresonant antennas (nearly 40 times larger SHG rate). The 2D SHG excitation maps obtained by scanning a single nanoantenna under the microscope objective focus are quantitatively accounted for in terms of intensity distribution and polarization response by finite element method simulations that model the exact experimental configuration (large NA, substrate, reflection mode, etc\ldots). Even the signal amplitudes expressed in photon/s are found in agreement when the transmission losses and detector efficiencies are considered. In a second step, we demonstrate that the SHG sources arising from the \(\chi_{\perp\perp\perp}\) local surface and the \(\gamma_{bulk}\) nonlocal bulk contributions can lead to distinct intensity distributions while imaging resonant nanostructures excited with a tightly focused beam. They are therefore separable provided that \(\chi_{\perp\parallel\parallel}\) is assumed to be zero, in agreement with the existing models for metals. We further indentify the dominant contribution to the nonlinear process: in contrast to gold nanostructures, the symmetry breaking induced by the surface is the major SHG source in aluminum nanoantennas. Hence, aluminum nanostructures can now be confidently modeled and used as building blocks to boost the nonlinear efficiency at the nanoscale. It includes the design of novel hybrid structures based on nonlinear and nanosized monocrystals embedded in plasmonic antennas and their optimization through quantitative (in terms of absolute count rates) simulations. The latter will allow answering the long-standing question of the relative efficiencies of plasmonic, dielectric and semiconducting materials as SHG generators at the nanoscale for potential integration. Their extension to sum frequency generation (SFG) is straightforward in the context of the developed approach and is indeed already underway. It opens new potentialities for ultrafast modulators with optimized designs, which are furthermore investigable with our current experimental setup.

\section*{Experimental}

Investigating nonlinear processes at the single particle level requires diffraction-limited spatial resolution, high sensitivity and low noise experimental setups. In order to reach these specifications, our setup has been designed as follows. The laser source is a femtosecond oscillator Ti:Sapphire working at a wavelength of 850 nm and delivering 100 fs pulses at a repetition rate of 80 MHz. For the measurements, the average input power is set at 245 \(\mu W\) to avoid damaging the material surface, except for Figure 1 for which it was reduced to 125 \(\mu W\). The focusing of the laser beam is provided by an immersion oil microscope objective (x100, numerical aperture \(NA=1.3\)), providing spatial resolutions of 330 and 165 nm at the fundamental and harmonic frequencies, respectively (\(\lambda/2NA\)). Nonlinear signals are collected in reflection mode and separated from the residual fundamental beam by a cold mirror. To overcome chromatic aberrations from the microscope objective, automated telescopes have been set in to overlap the focusing planes at the excitation and collection wavelengths with the surface of the sample. The latter is mounted on a set of closed-loop piezoelectric stages for lateral scanning with nanometer accuracy under the focused beam. Coupled with a tracking algorithm, the full computer control of the setup provides an automated optimization of the measured signals before any set of measurements, yielding extremely robust and reproducible results. The collected light is then spectrally selected by a monochromator with a weakly dispersive grating (150 g/mm) and analyzed through a polarizing beam-splitter coupled to two SPCMs. An ultra-fast acquisition card is synchronized with the laser impulsions to discriminate between the correlated SHG signal and random noise, reducing dark counts contributions to a few events/s.

Individual aluminum nanostructures are fabricated on a glass substrate by standard electron beam lithography. A microscope cover slip adapted to our microscope objective is cleaned in acetone before being prepared by immersion in nitric acid for 30 s and rinsed in deionized water. An oxygen plasma cleaning is then performed for 2 minutes to remove residual organic contamination of the surface. The sample is subsequently heated at 180\(^{\circ}\)C for 5 minutes before a PMMA 50kDa/PMMA 950kDa bi-layer resist is spin coated onto its surface. Each layer is baked at 80\(^{\circ}\)C for 5 minutes. An additional 10 nm thick layer of aluminum is deposited on the resist by electron-gun evaporation to help evacuating the residual charges during lithography. After lithography, the aluminum layer is dissolved in a strong base solution. The resist is developed in MIBK:IPA and fixed by IPA rinsing. A final 35 nm thick aluminum layer is deposited onto the sample using electron-gun evaporation, before N-methyl-2-pyrrolidone (NMP) lift-off is performed at 80\(^{\circ}\)C. Lateral dimensions of the antennas (100 nm in width and 125 nm to 525 nm in length) are measured using a scanning electron microscope (SEM), while the 35 nm thickness is confirmed by scanning the tip of a near-field optical microscope\cite{Chevalier2006} implemented on the optical setup and used in topographic mode (see insets of Figure 1).

\begin{acknowledgement}

The authors acknowledge the financial support of the Agence Nationale de la Recherche (ANR) [Grant ANR-14-CE26-
0001-01-TWIN], the Universit\'e Grenoble Alpes [Chaire IUA] and the Labex LANEF [PhD program].

\end{acknowledgement}

%%%%\bibliography{achemso-demo}

\end{document}